\documentclass[10pt,twocolumn,letterpaper]{article}

\usepackage[pagenumbers]{iccv} % To force page numbers, e.g. for an arXiv version

\usepackage{times}
\usepackage{epsfig}
\usepackage{graphicx}
\usepackage{amsmath}
\usepackage{amssymb}

\usepackage[export]{adjustbox}

\usepackage{booktabs}

\usepackage{float}

\usepackage{stfloats} % added by author2

\usepackage{dsfont}
\usepackage{subcaption}
\usepackage{caption}
\usepackage{nicefrac}
\usepackage{mathrsfs}
\usepackage{xcolor}
\usepackage{enumitem}
\usepackage{colortbl}
\usepackage{multirow}

\usepackage{marvosym}

\usepackage{pifont}

\definecolor{sh_gray}{rgb}{0.84,0.84,0.84}
\definecolor{sh_gray2}{rgb}{1,0.89,0.75}
\definecolor{color3}{rgb}{0.95,0.95,0.95}
\definecolor{color4}{rgb}{0.94,0.94,1}
\definecolor{color5}{rgb}{1,0.96,0.88}

\def\xnet{GuideSR\xspace}

\newif\ifdraft
\draftfalse
\ifdraft
\newcommand{\jwc}[1]{{\color{orange}[\textbf{Jian} #1]}}
\newcommand{\smc}[1]{{\color{red}[\textbf{Sizhuo} #1]}}
\newcommand{\yfc}[1]{{\color{blue}[\textbf{Yufei} #1]}}

\newcommand{\sm}[1]{{\color{red}#1}}

\else
\newcommand{\jwc}[1]{}
\newcommand{\smc}[1]{}
\newcommand{\yfc}[1]{}

\newcommand{\sm}[1]{{\color{black}#1}}

\fi

\usepackage[pagebackref=true,breaklinks=true,letterpaper=true,colorlinks,bookmarks=false]{hyperref}

\definecolor{iccvblue}{rgb}{0.21,0.49,0.74}
\def\paperID{} % *** Enter the Paper ID here
\def\confName{ICCV}
\def\confYear{2025}

\newcommand\blfootnote[1]{%
  \begingroup
  \renewcommand\thefootnote{}\footnote{#1}%
  \addtocounter{footnote}{-1}%
  \endgroup
}

\begin{document}

%%%%%%%%% TITLE
\title{GuideSR: Rethinking Guidance for One-Step High-Fidelity Diffusion-Based Super-Resolution}
\author{Aditya Arora$^{1\;\star}$ \quad Zhengzhong Tu$^{2\;\dagger}$ \quad Yufei Wang$^{3\;\dagger}$ \\
Ruizheng Bai$^{2}$ \quad Jian Wang$^{3\;\ddagger}$ \quad Sizhuo Ma$^{3\;\ddagger}$ \\
$^1$ TU Darmstadt \quad $^2$ Texas A\&M University \quad $^3$ Snap Inc. \\
{\tt\small aditya.arora@tu-darmstadt.de \{tzz,rzbai\}@tamu.edu \{ywang25,jwang4,sma\}@snapchat.com}
}

\maketitle
\blfootnote{$\star$ Work done during internship at Snap Inc.}
\blfootnote{$\dagger$ Equal Contribution $\ddagger$ Co-corresponding Author}

%%%%%%%%% ABSTRACT
\begin{abstract}
In this paper, we propose \xnet, a novel single-step diffusion-based image super-resolution (SR) model specifically designed to enhance image fidelity. Existing diffusion-based SR approaches typically adapt pre-trained generative models to image restoration tasks by adding extra conditioning on a VAE-downsampled representation of the degraded input, which often compromises structural fidelity. \xnet addresses this limitation by introducing a dual-branch architecture comprising: (1) a Guidance Branch that preserves high-fidelity structures from the original-resolution degraded input, and (2) a Diffusion Branch, which a pre-trained latent diffusion model to enhance perceptual quality. Unlike conventional conditioning mechanisms, our Guidance Branch features a tailored structure for image restoration tasks, combining Full Resolution Blocks (FRBs) with channel attention and an Image Guidance Network (IGN) with guided attention. By embedding detailed structural information directly into the restoration pipeline, \xnet produces sharper and more visually consistent results. Extensive experiments on benchmark datasets demonstrate that \xnet achieves state-of-the-art performance while maintaining the low computational cost of single-step approaches, with up to 1.39dB PSNR gain on challenging real-world datasets. Our approach consistently outperforms existing methods across various reference-based metrics including PSNR, SSIM, LPIPS, DISTS and FID,
further representing a practical advancement for real-world image restoration.

\end{abstract}

%%%%%%%%% BODY TEXT
\section{Introduction}

\begin{figure}
    \centering
    \includegraphics[width=\linewidth]{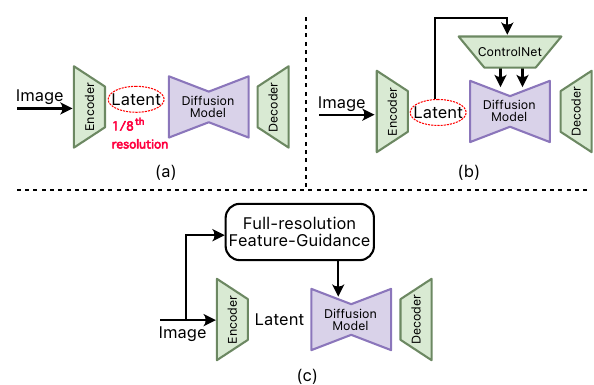}
    \caption{\textbf{Architecture comparison of diffusion-based super-resolution approaches}. (a) Standard diffusion process (e.g., OSEDiff~\cite{osediff}) processes latent representations directly;
    (b) Controller-based methods (e.g., DiffBIR~\cite{diffbir}, SeeSR~\cite{seesr}, StableSR~\cite{stablesr}) employ conditional mechanisms to guide the diffusion process; (c) Our proposed GuideSR introduces a dual-branch architecture with full-resolution feature guidance, preserving high-frequency details from the original input while leveraging the generative capabilities of diffusion models. This approach addresses the key limitation of existing methods the loss of structural fidelity due to VAE encoding of degraded inputs.}
    \label{fig:teaser}
\end{figure}

Image super-resolution (SR) aims to reconstruct high-resolution (HR) images from their low-resolution (LR) counterparts~\cite{dong2014learning, rcan,liang2021swinir, zhang2022efficient, chen2023activating, srgan, wang2018esrgan, liang2022details, zhang2021designing}. Beyond merely refining visible structures, SR seeks to recover details lost due to resolution degradation. 
This characteristic inherently renders SR an ill-posed problem, as multiple plausible HR images may correspond to the same LR input, necessitating models to accurately infer missing high-frequency details while maintaining fidelity to the original, often degraded content. Traditional SR methods have addressed this challenge using handcrafted priors~\cite{freedman2011image, glasner2009super}, regression-based deep neural networks~\cite{dong2014learning, kim2016accurate}, as well as generative adversarial networks~\cite{ledig2017photo, wang2018esrgan}.

In recent years, diffusion models have emerged as powerful frameworks for high-quality image generation~\cite{ho2020denoising, rombach2022high, saharia2022photorealistic, nichol2021improved}. These models operate through an iterative denoising process that gradually transforms random noise into coherent images, enabling high-quality image generation with unprecedented detail and diversity. It has been shown that such generative capabilities can also serve as a strong prior for recovering missing details in SR~\cite{sr3, stablesr} and other low-level vision tasks~\cite{diffbir, kawar2022denoising,li2024light}. However, early diffusion-based methods typically require numerous denoising steps (e.g., 50-200) to recover a single image, significantly limiting their practicality. Recent advancements have proposed single-step restoration~\cite{sinsr, osediff}, offering a promising direction for efficient and realistic super-resolution under real-world degradations.

While single-step diffusion SR methods~\cite{sinsr, osediff} greatly improve inference efficiency, they still struggle to preserve structural fidelity alongside realistic texture generation~\cite{pasd,wan2024clearsr}. We hypothesize that this limitation arises from the way existing methods condition their diffusion processes on LR inputs. As illustrated in Figure~\ref{fig:teaser},  current approaches typically encode the LR input to a latent space via a pretrained variational autoencoder (VAE). Then they either directly feed this LR latent into the denoising UNet (\eg OSEDiff~\cite{osediff}), or condition on the LR latent through a controller mechanism~\cite{zhang2023adding} (\eg DiffBIR~\cite{diffbir}, SeeSR~\cite{seesr}, and StableSR~\cite{stablesr}). Yet, since the VAE often employs aggressive downsampling with a high compression ratio (\eg 8x~\cite{rombach2022high}), it inevitably leads to the loss of high-frequency spatial details. Furthermore, because VAEs are primarily trained on high-quality images~\cite{osediff}, their encoding process can degrade structural integrity when applied to lower-quality LR inputs. As a result, these approaches frequently exhibit suboptimal performance, particularly in reconstructing complex textures and fine patterns, failing to effectively balance detail hallucination and faithful reconstruction.

To address these limitations, we rethink the guidance mechanism design by taking an \emph{SR-first} approach for diffusion-based SR modeling. We introduce \xnet, a novel single-step diffusion-based SR method with a dual-branch architecture specifically designed to overcome the structural fidelity challenges on which most previous approaches struggle. As illustrated in Figure~\ref{fig:teaser}, our architecture consists of two complementary branches: \textbf{(1)} a \underline{Guidance Branch} operating at full resolution to preserve structural details, and \textbf{(2)} a \underline{Diffusion Branch}, which leverages the generative capabilities of a pre-trained latent diffusion model to enhance perceptual quality. Departing from traditional reliance on VAEs and controllers \cite{zhang2023adding, rombach2022high}, the Guidance Branch is specifically tailored to restoration, which directly processes the \emph{full-resolution} low-quality input, bypassing the VAE latent encoding that causes detail loss in previous methods. This branch consists of Full Resolution Blocks (FRB) with Channel Attention mechanisms arranged in a residual-in-residual structure, along with an Image Guidance Network (IGN) that uses guided attention to preserve high-frequency details. Together, these components ensure effective feature extraction for high-fidelity reconstruction. Operating at the original image resolution allows this branch to preserve fine textures and structural information that would otherwise be lost in latent space processing.
The main contributions of this work are:
\begin{itemize}[leftmargin=*]
    \item We introduce GuideSR, a novel single-step diffusion-based SR framework featuring a dual-branch architecture that effectively balances structural fidelity with perceptual quality, addressing a fundamental limitation in existing diffusion-based restoration methods.
    \item We design innovative Full Resolution Blocks and Image Guidance Network that adaptively refines full-resolution features, ensuring high-frequency details are preserved throughout the restoration process.
    \item We demonstrate, through extensive experiments, that our approach achieves state-of-the-art performance on multiple benchmarks, significantly surpassing existing methods in terms of fidelity and perceptual quality while maintaining computational efficiency.
\end{itemize}

\begin{figure*}
    \centering
    \includegraphics[width=0.98\linewidth]{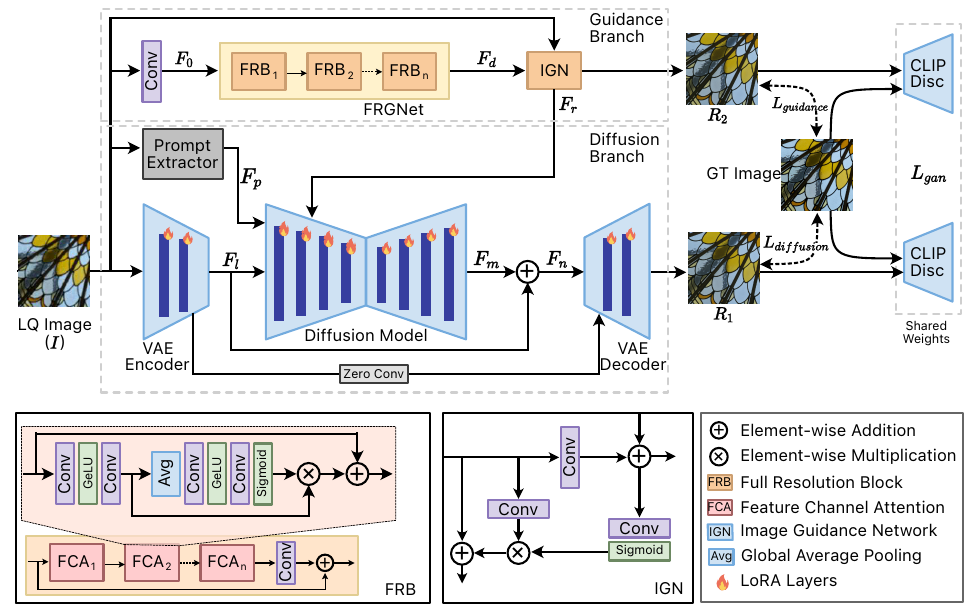}
    \caption{\textbf{Overview of \xnet architecture.} Our method introduces a dual-branch architecture where the Guidance Branch (top) processes full-resolution input to preserve high-frequency details, while the Diffusion Branch (bottom) operates in the latent space for enhanced perceptual quality. The Guidance Branch applies a series of Full Resolution Blocks (FRBs) and an Image Guidance Network (IGN) to output a refined image $R_2$, which are enabled by the channel attention mechanisms and feature refinement operations detailed in the bottom. The Diffusion Branch employs a LoRA-finetuned diffusion model to produce the final output $R_1$, where the features from the Guidance Branch are adaptively refined and integrated into the denoise U-Net. Both branches are supervised through discriminators with shared weights during training.  }
    \label{fig:architecture}
\end{figure*}

\section{Related Work}

\subsection{Efficient Diffusion Models}
While Denoising Diffusion Probabilistic Models (DDPMs)~\cite{ho2020denoising}  demonstrate exceptional image generation capabilities, they require 1000 training steps and typically 50 inference steps, which is prohibitive for many real-world applications. Latent Diffusion Models (LDM) and Stable Diffusion (SD)~\cite{rombach2022high} enables generation of high-resolution images through a more computationally feasible latent space, yet they continue to depend on an iterative denoising process with billions of parameters. To mitigate this, faster samplers~\cite{song2020denoising,lu2022dpm,dockhorn2022genie} have been developed to reduce the number of denoising steps. Recently, step distillation methods~\cite{meng2023distillation,salimans2022progressive} have emerged, distilling a pretrained diffusion model to significantly fewer steps. Notably, SD Turbo introduced Adversarial Diffusion Distillation (ADD)~\cite{sauer2024adversarial}, facilitating single-step image synthesis. Beyond reducing the number of steps, techniques such as architecture optimization~\cite{kim2024bk,hu2024snapgen} and quantization~\cite{li2023q,huang2024tfmq} have been implemented to further improve the efficiency of diffusion models, making photorealistic image generation feasible on smartphones.

\subsection{Diffusion Models for Image Super Resolution}

Diffusion models have shown promising results in image super-resolution (SR). 
Early works like SR3 \cite{sr3} adapted DDPM frameworks to SR tasks, demonstrating significant improvements in perceptual quality. StableSR \cite{stablesr} was the first to leverage prior knowledge in a pre-trained text-to-image diffusion model. DiffBIR \cite{diffbir} introduced a two-stage model for degradation removal and realistic image reconstruction. SeeSR \cite{seesr} utilized semantic prompts to generate detailed and semantically accurate results. CoDi~\cite{mei2024codi} presented a conditional diffusion distillation method to accelerate diffusion sampling. PASD \cite{pasd} proposed pixel-aware cross attention to effectively inject the pixel-level information into the diffusion model. ResShift \cite{resshift} introduced a residual shift mechanism that significantly reduces the number of denoise steps. SinSR \cite{sinsr} further simplifies ResShift to a single-step model via  consistency preserving distillation. OSEDiff \cite{osediff} achieves single-step inference through LoRA-finetuning a pre-trained Stable Diffusion model with variational score distillation.  Despite these advances, preserving structural information remains a challenge for diffusion-based SR models. Our work addresses this challenge by explicitly integrating structure information from the full-resolution LR input using a Guidance Branch specifically designed for image restoration, achieving higher fidelity compared to previous techniques.

\section{Methodology}

Our goal is to develop a single-step SR model that leverages the generative capabilities of diffusion priors without compromising the structural integrity of input images. 
To achieve this, we introduce GuideSR, a dual-branch architecture consisting of a Guidance Branch for efficient full-resolution feature- and image-level guidance and an enhanced Diffusion Branch for high-quality details synthesis. These branches work in tandem to refine degraded inputs while maintaining computational efficiency. In the following sections, we detail the architecture and operation of each branch (Sections \ref{subsec:guidance_branch} and \ref{subsec:diffusion_branch}), and describe our training strategy (Section \ref{subsec:training_strategy}).

\subsection{Guidance Branch}
\label{subsec:guidance_branch}

Existing diffusion-based SR models rely on VAE and ControlNet to condition a diffusion model on a degraded input image, which effectively capture the image context but struggle to retain fine-grained textures that are crucial for high-quality image restoration. To address this limitation, our Guidance Branch operates at the original image resolution, preserving structural and textural integrity.

Given a degraded image $I \in \mathbb{R}^{H\times W\times 3}$, we first apply a convolutional layer to extract low-level feature embeddings:
\begin{equation}
F_0 = \text{Conv}(I), \quad F_0 \in \mathbb{R}^{H\times W\times C},
\end{equation}
where $H \times W$ represents the spatial dimensions, and $C$ denotes the number of feature channels. These shallow features $F_0$ are then passed through multiple Full Resolution Blocks (FRBs) to obtain deep features:
\begin{equation}
\scalebox{0.85}{$
\begin{aligned}
F_d &= \text{FRGNet}(F_0) \\
&= \text{FRB}_n \circ \ldots \circ \text{FRB}_{2} \circ \text{FRB}_1(F_0), \quad F_d \in \mathbb{R}^{H\times W\times 2C},
\end{aligned}
$}
\end{equation}
where $\circ$ denotes function composition and $\text{FRB}_i$ represents the $i$-th Full Resolution Block in the sequence.

As shown in Figure~\ref{fig:architecture}, each FRB consists of multiple Feature Channel Attention (FCA) blocks and skip connection arranged in a residual-in-residual structure. The detailed structure of an FRB, shown in the bottom-left of Figure~\ref{fig:architecture}, includes a standard channel attention pathway. Each FRB implements a residual connection where the input is added to a feature-recalibrated version of itself. The recalibration applies convolution, GeLU activation, average pooling, another convolution with sigmoid activation, and element-wise multiplication with the input, followed by a final convolution. This design enables effective feature representation by adaptively emphasizing crucial image regions while suppressing less relevant details.

To further refine the features, we employ an Image Guidance Network (IGN) which leverages guided attention to enhance structural integrity, as shown in Figure~\ref{fig:architecture} (bottom-center). The IGN applies a residual connection where the input features $F_d$ are added to an attention-modulated version of themselves. The attention mechanism involves applying convolutions and sigmoid activation to generate attention maps, which are then multiplied with another convolutional projection of the input features.

The IGN outputs two key components: a refined residual image $R_2$, which contributes to loss computation during training, and enriched feature representations $F_r$ containing structural information. The refined residual image is obtained by adding a convolutional projection of the refined features to the original input image: $R_2 = \text{Conv}(F_r) + I$.

To align the spatial resolution of the refined features with the diffusion model's UNet-Encoder features, we apply pixel-unshuffle operations for downsampling:
\begin{equation}
F'_r = \text{PixelUnshuffle}(F_r, s) \in \mathbb{R}^{\frac{H}{s} \times \frac{W}{s} \times s^2 \cdot 2C},
\end{equation}
where $s$ is the downsampling factor. The pixel-unshuffle operation rearranges the spatial dimensions of a tensor while preserving all values, making it ideal for the SR task.

The downsampled features $F'_r$ are then concatenated with UNet encoder outputs to enrich hierarchical feature representations at multiple scales, as illustrated by the vertical connections in Figure~\ref{fig:architecture}. This multi-scale integration allows structural information to influence the diffusion process at various resolutions, preserving details that might otherwise be lost.

\begin{table*}[!t]
\begin{center}
% \caption{\small \underline{\textbf{Image Super Resolution}} results on DIV2K-Val dataset.}
\caption{\small \underline{\textbf{Quantitative comparison of diffusion-based super-resolution methods}} across DIV2K-Val, DRealSR, and RealSR datasets. GuideSR consistently outperforms both multi-step methods (requiring 15-200 steps) and recent one-step approaches \sm{across all reference-based metrics (PSNR/SSIM/LPIPS/DISTS/FID). Our method achieves significant improvements on the challenging real-world DRealSR dataset with a \textbf{1.39dB PSNR gain} over the best multi-step method (ResShift) while maintaining the efficiency of single-step inference. } 
% across key fidelity (PSNR/SSIM) and perceptual (LPIPS/DISTS/FID) metrics. 
% Our method achieves significant improvements in the challenging real-world DRealSR dataset with a \textbf{1.22dB PSNR gain} over the best multi-step method (ResShift) while maintaining superior perceptual quality with the lowest LPIPS and DISTS scores across all datasets. 
% These results demonstrate the effectiveness of our dual-branch architecture in balancing structural fidelity with perceptual quality while maintaining the efficiency of single-step inference. 
Bold values indicate the best results for each metric.}
% \vspace{-2mm}
\label{table:sota_table}
% \vspace{-2mm}
% \setlength{\tabcolsep}{9.5pt}
\setlength{\tabcolsep}{5.6pt}
\scalebox{0.85}{
\begin{tabular}{l | l c | c c c c c | c c c c}
\toprule[0.15em]
 \textbf{Dataset} & \textbf{Method} & Steps & PSNR~$\textcolor{black}{\uparrow}$ & SSIM~$\textcolor{black}{\uparrow}$ & LPIPS~$\textcolor{black}{\downarrow}$ & DISTS~$\textcolor{black}{\downarrow}$ & FID~$\textcolor{black}{\downarrow}$ & NIQE~$\textcolor{black}{\downarrow}$ & MUSIQ~$\textcolor{black}{\uparrow}$ & MANIQA~$\textcolor{black}{\uparrow}$ & CLIPIQA~$\textcolor{black}{\uparrow}$ \\
\midrule[0.15em]
\multirow{8}{*}{DIV2K} & StableSR \cite{stablesr} & 200 & 23.26 & 0.5726 & 0.3113 & 0.2048 & 24.44 & 4.7581 & 65.92 & 0.6192 & 0.6771 \\
& DiffBIR \cite{diffbir}      & 50  & 23.64 & 0.5647 & 0.3524 & 0.2128 & 30.72 & 4.7042 & 65.81 & 0.6210 & 0.6704 \\
& SeeSR \cite{seesr}           & 50  & 23.68 & 0.6043 & 0.3194 & 0.1968 & 25.90 & 4.8102 & 68.67 & 0.6240 & \textbf{0.6936} \\
& PASD \cite{pasd}          & 20  & 23.14 & 0.5505 & 0.3571 & 0.2207 & 29.20 & \textbf{4.3617} & \textbf{68.95} & \textbf{0.6483} & 0.6788 \\
& ResShift \cite{resshift}    & 15  & 24.65 & 0.6181 & 0.3349 & 0.2213 & 36.11 & 6.8212 & 61.09 & 0.5454 & 0.6071 \\
\cmidrule{2-12}
& SinSR \cite{sinsr}         & 1   & 24.41 & 0.6018 & 0.3240 & 0.2066 & 35.57 & 6.0159 & 62.82 & 0.5386 & 0.6471 \\
& OSEDiff \cite{osediff}           & 1   & 23.72 & 0.6108 & 0.2941 & 0.1976 & 26.32 & 4.7097 & 67.97 & 0.6148 & 0.6683 \\
\cmidrule{2-12}
% & \textbf{\xnet} & \textbf{1} & \textbf{24.76} & \textbf{0.6333} & \textbf{0.2653} & \textbf{0.1879} & \textbf{21.04} & 5.9288 & 63.96 & 0.5679 & 0.5893 \\
& \textbf{\xnet} & \textbf{1} & \textbf{24.76} & \textbf{0.6333} & \textbf{0.2653} & \textbf{0.1879} & \textbf{21.04} & 5.9273 & 63.97 & 0.5679 & 0.5840 \\
\bottomrule[0.1em]
% \end{tabular}}
% \end{center}\vspace{-1.5em}
% \end{table*}
% \begin{table*}[!t]
% \begin{center}
% \caption{\small \underline{\textbf{Image Super Resolution}} results on DRealSR dataset.}
% \label{table:drealsr}
% \vspace{-2mm}
% \setlength{\tabcolsep}{6pt}
% % \setlength{\tabcolsep}{9.5pt}
% \scalebox{0.85}{
% \begin{tabular}{l c | c c c c c | c c c c}
% \toprule[0.15em]
%  \textbf{Method} & Steps & PSNR~$\textcolor{black}{\uparrow}$ & SSIM~$\textcolor{black}{\uparrow}$ & LPIPS~$\textcolor{black}{\downarrow}$ & DISTS~$\textcolor{black}{\downarrow}$ & FID~$\textcolor{black}{\downarrow}$ & NIQE~$\textcolor{black}{\downarrow}$ & MUSIQ~$\textcolor{black}{\uparrow}$ & MANIQA~$\textcolor{black}{\uparrow}$ & CLIPIQA~$\textcolor{black}{\uparrow}$ \\
% \midrule[0.15em]
\multirow{8}{*}{DRealSR} & StableSR \cite{stablesr} & 200 & 28.03 & 0.7536 & 0.3284 & 0.2269 & 148.98 & 6.5239 & 58.51  & 0.5601  & 0.6356 \\
& DiffBIR \cite{diffbir} & 50 & 26.71 & 0.6571 & 0.4557 & 0.2748 & 166.79 & 6.3124 & 61.07  & 0.5930  & 0.6395 \\
& SeeSR \cite{seesr} & 50 & 28.17 & 0.7691 & 0.3189 & 0.2315 & 147.39 & 6.3967 & \textbf{64.93}  & 0.6042  & 0.6804 \\
& PASD \cite{pasd} & 20 & 27.36 & 0.7073 & 0.3760 & 0.2531 & 156.13 & \textbf{5.5474} & 64.87  & \textbf{0.6169}  & 0.6808 \\
& ResShift \cite{resshift} & 15 & 28.46 & 0.7673 & 0.4006 & 0.2656 & 172.26 & 8.1249 & 50.60  & 0.4586  & 0.5342 \\
\cmidrule{2-12}
& SinSR \cite{sinsr} & 1 & 28.36 & 0.7515 & 0.3665 & 0.2485 & 170.57 & 6.9907 & 55.33  & 0.4884  & 0.6383 \\
& OSEDiff \cite{osediff} & 1 & 27.92 & 0.7835 & 0.2968 & 0.2165 & 135.30 & 6.4902 & 64.65  & 0.5899  & \textbf{0.6963} \\
\cmidrule{2-12}
% & \textbf{\xnet} & \textbf{1} & \textbf{29.68} & \textbf{0.7995} & \textbf{0.2709} & \textbf{0.2017} & \textbf{124.84} & 7.5282 & 58.96 & 0.5298 & 0.5792 \\
& \textbf{\xnet} & \textbf{1} & \textbf{29.85} & \textbf{0.8078} & \textbf{0.2640} & \textbf{0.1960} & \textbf{122.06} & 7.7500 & 57.14 & 0.5230 & 0.5762 \\
% \bottomrule[0.1em]
% \end{tabular}}
% \end{center}\vspace{-1.5em}
% \end{table*}
% \begin{table*}[!t]
% \begin{center}
% \caption{\small \underline{\textbf{Image Super Resolution}} results on RealSR dataset.}
% \label{table:drealsr}
% \vspace{-2mm}
% \setlength{\tabcolsep}{9.5pt}
% \setlength{\tabcolsep}{6pt}
% \scalebox{0.85}{
% \begin{tabular}{l c | c c c c c | c c c c}
% \toprule[0.15em]
 % \textbf{Method} & Steps & PSNR~$\textcolor{black}{\uparrow}$ & SSIM~$\textcolor{black}{\uparrow}$ & LPIPS~$\textcolor{black}{\downarrow}$ & DISTS~$\textcolor{black}{\downarrow}$ & FID~$\textcolor{black}{\downarrow}$ & NIQE~$\textcolor{black}{\downarrow}$ & MUSIQ~$\textcolor{black}{\uparrow}$ & MANIQA~$\textcolor{black}{\uparrow}$ & CLIPIQA~$\textcolor{black}{\uparrow}$ \\
\midrule[0.15em]
\multirow{8}{*}{RealSR} & StableSR \cite{stablesr} & 200 & 24.70 & 0.7085 & 0.3018 & 0.2288 & 128.51 & 5.9122 & 65.78  & 0.6221  & 0.6178 \\
& DiffBIR \cite{diffbir} & 50 & 24.75 & 0.6567 & 0.3636 & 0.2312 & 128.99 & 5.5346 & 64.98  & 0.6246  & 0.6463 \\
& SeeSR \cite{seesr} & 50 & 25.18 & 0.7216 & 0.3009 & 0.2223 & 125.55 & 5.4081 & \textbf{69.77}  & 0.6442  & 0.6612 \\
& PASD \cite{pasd} & 20 & 25.21 & 0.6798 & 0.3380 & 0.2260 & 124.29 & \textbf{5.4137} & 68.75  & \textbf{0.6487}  & 0.6620 \\
& ResShift \cite{resshift} & 15 & 26.31 & 0.7421 & 0.3460 & 0.2498 & 141.71 & 7.2635 & 58.43  & 0.5285  & 0.5444 \\
\cmidrule{2-12}
& SinSR \cite{sinsr} & 1 & 26.28 & 0.7347 & 0.3188 & 0.2353 & 135.93 & 6.2872 & 60.80  & 0.5385  & 0.6122  \\
& OSEDiff \cite{osediff} & 1 & 25.15 & 0.7341 & 0.2921 & 0.2128 & 123.49 & 5.6476 & 69.09  & 0.6326  & \textbf{0.6693}  \\
\cmidrule{2-12}
% & \textbf{\xnet} & \textbf{1} & \textbf{26.97} & \textbf{0.7633} & \textbf{0.2432} & \textbf{0.1901} & \textbf{97.49} & 6.5131 & 64.15 & 0.5842 & 0.5693 \\
& \textbf{\xnet} & \textbf{1} & \textbf{27.08} & \textbf{0.7681} & \textbf{0.2407} & \textbf{0.1878} & \textbf{96.83} & 6.7647 & 62.20 & 0.5716 & 0.5482 \\
\bottomrule[0.1em]
\end{tabular}}
\end{center}
% \vspace{-6mm}
\end{table*}

\subsection{Diffusion Branch}
\label{subsec:diffusion_branch}

{We use a pretrained latent diffusion model as our generative prior, as shown in Figure~\ref{fig:architecture} (bottom branch).
Starting from the high-resolution input, the VAE Encoder progressively reduces spatial dimensions while expanding channel capacity. The encoder transforms the input image $I$ into latent features $F_l = \text{VAE}_{\text{Encoder}}(I) \in \mathbb{R}^{\frac{H}{8} \times \frac{W}{8} \times 4C}$.

Additionally, a text prompt is extracted from the input image using a dedicated prompt extraction module. This produces a text embedding vector $F_p = \text{PromptExtractor}(I)$, allowing the diffusion model to leverage text-conditional generation capabilities and guide the restoration process with semantic understanding.

The central UNet structure processes these latent features with the integrated guidance from both the text prompt and the downsampled Guidance Branch features. 
While the pretrained diffusion model is conditioned on the timestep t, we choose a fixed $t_f$ for the one-step model: 
\begin{equation}
F_m = \text{UNet}(F_l, t_f, F_p, \{F'_r\}) \in \mathbb{R}^{\frac{H}{8} \times \frac{W}{8} \times 4C},
\end{equation}
where $\{F'_r\}$ represents the set of multi-scale features from the Guidance Branch. Notice that a long-skip connection is added such that the UNet only predicts the \emph{residual} latent, with the final latent features computed as:
\begin{equation}
    F_n = F_m + F_l.
\end{equation}
This skip connection ensures that low-frequency information from the original input is preserved throughout the diffusion process, allowing the network to focus on generating high-frequency details while maintaining overall image structure. The output of the UNet is then decoded through the VAE Decoder to produce the result $R_1 = \text{VAEDecoder}(F_n) \in \mathbb{R}^{H\times W\times 3}$.

During training, LoRA layers~\cite{lora} are added for parameter-efficient finetuning. For a weight matrix $W \in \mathbb{R}^{d\times k}$ in the original network, LoRA parameterizes the update as a low-rank decomposition $\Delta W = BA$, where $B \in \mathbb{R}^{d\times r}$, $A \in \mathbb{R}^{r\times k}$, and $r \ll \min(d, k)$ are the ranks. Specifically, we employ LoRA ranks of $r = 8$ for UNet layers and $r = 4$ for the VAE. We also add skip-connections with Zero-Convs between the VAE Encoder and Decoder~\cite{parmar2024one} to mitigate the detail loss due to VAE in the Diffusion Branch.

\subsection{Training Strategy}
\label{subsec:training_strategy}

Previous work has shown that adversarial training help reduce the number of steps of diffusion models significantly~\cite{sauer2024adversarial,xu2024ufogen,xiao2021tackling,lin2024sdxl}. We utilizes dual discriminators (shown in blue in Figure~\ref{fig:architecture}) that evaluate both the Guidance Branch output $R_2$ and the Diffusion Branch output $R_1$. These discriminators share weights to ensure consistency in the adversarial training signal. With this adversarial training, we are able to leverage the pretrained diffusion prior and train our one-step restoration model by computing the following loss directly from the restored image and the ground truth:
\begin{equation}
\mathcal{L}_{B} = \lambda_{1} \cdot \mathcal{L}_\text{MSE} + \lambda_{2} \cdot \mathcal{L}_\text{LPIPS} + \lambda_{3} \cdot \mathcal{L}_\text{GAN},
\end{equation}
where the individual loss terms are defined as:
\begin{equation}
\begin{split}
&\mathcal{L}_\text{MSE}(R, Y) = \frac{1}{HW}\sum_{i=1}^{H}\sum_{j=1}^{W}||R_{i,j} - Y_{i,j}||_2^2\\
&\mathcal{L}_\text{LPIPS}(R, Y) = ||\varPhi(R) - \varPhi(Y)||_2^2\\
&\mathcal{L}_\text{GAN}(R) = -\mathbb{E}[\log(D(R))]\\
\end{split}
\end{equation}
where $R$ is the restored image, $Y$ is the ground truth, $\varPhi$ represents the LPIPS network, and $D$ is the discriminator network. The weighting coefficients are set as $\lambda_\text{1} = 1.0$, $\lambda_\text{2} = 5.0$, and $\lambda_\text{3} = 0.5$.

While we use the Diffusion Branch as the final output during inference, computing the loss on the Guidance Branch helps stabilize training.
The final loss combines both branches with different weights:
\begin{equation}
\mathcal{L}_\text{final} = \lambda_d \cdot \mathcal{L}_B(Y, R_1) + \lambda_g \cdot \mathcal{L}_B(Y, R_2),
\end{equation}
where $Y$ is the ground truth image, $R_1$ is the output from the Diffusion Branch, and $R_2$ is the output from the Guidance Branch, with $\lambda_d = 0.9$ and $\lambda_g = 0.1$.

\section{Results}

\begin{figure*}[t!]
    \centering
    \begin{subfigure}[t]{0.26\textwidth}
        \centering
        \includegraphics[width=\textwidth,valign=m]{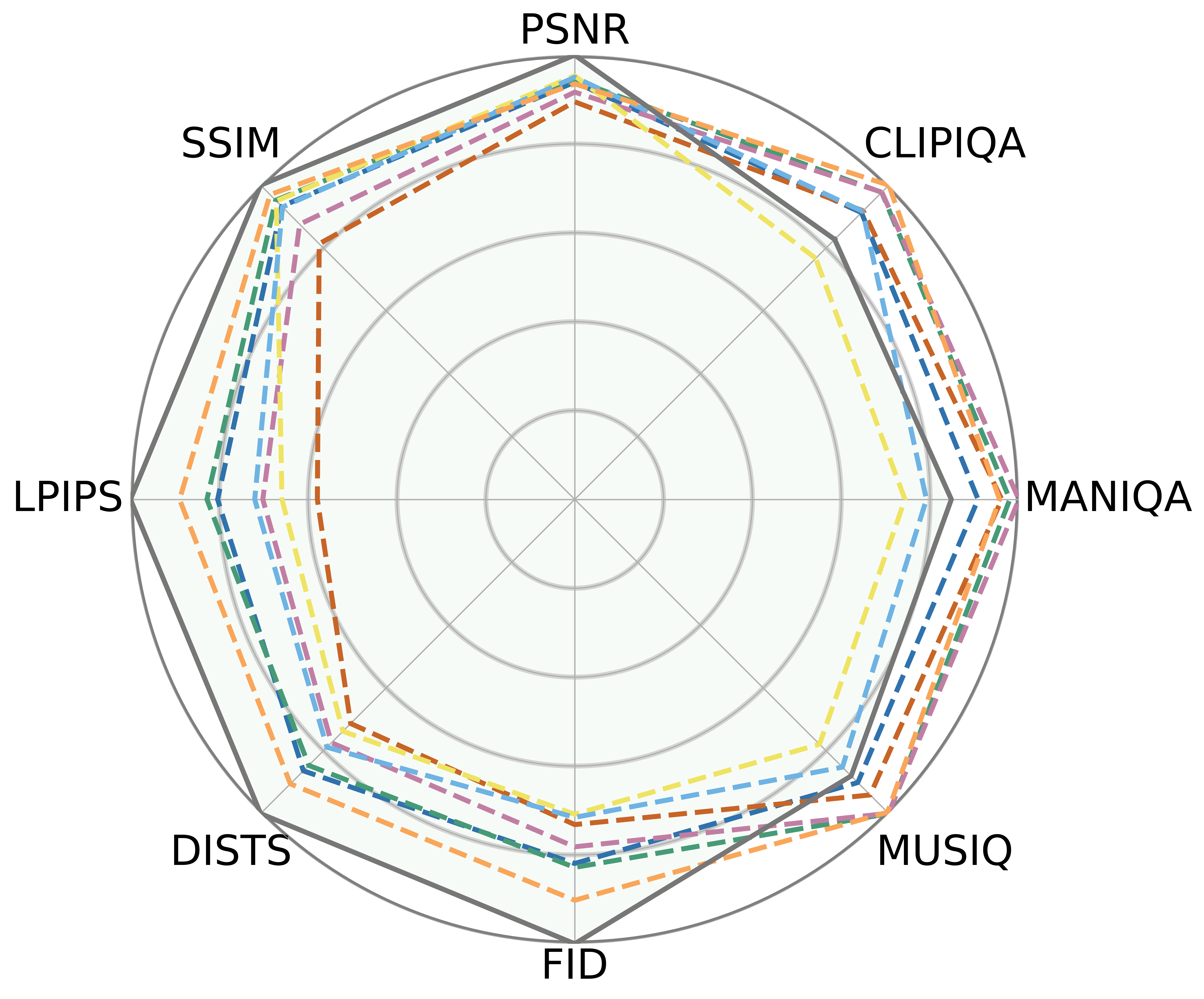}
        \caption{RealSR}
    \end{subfigure}%
    ~ 
    \begin{subfigure}[t]{0.26\textwidth}
        \centering
        \includegraphics[width=\textwidth,valign=m]{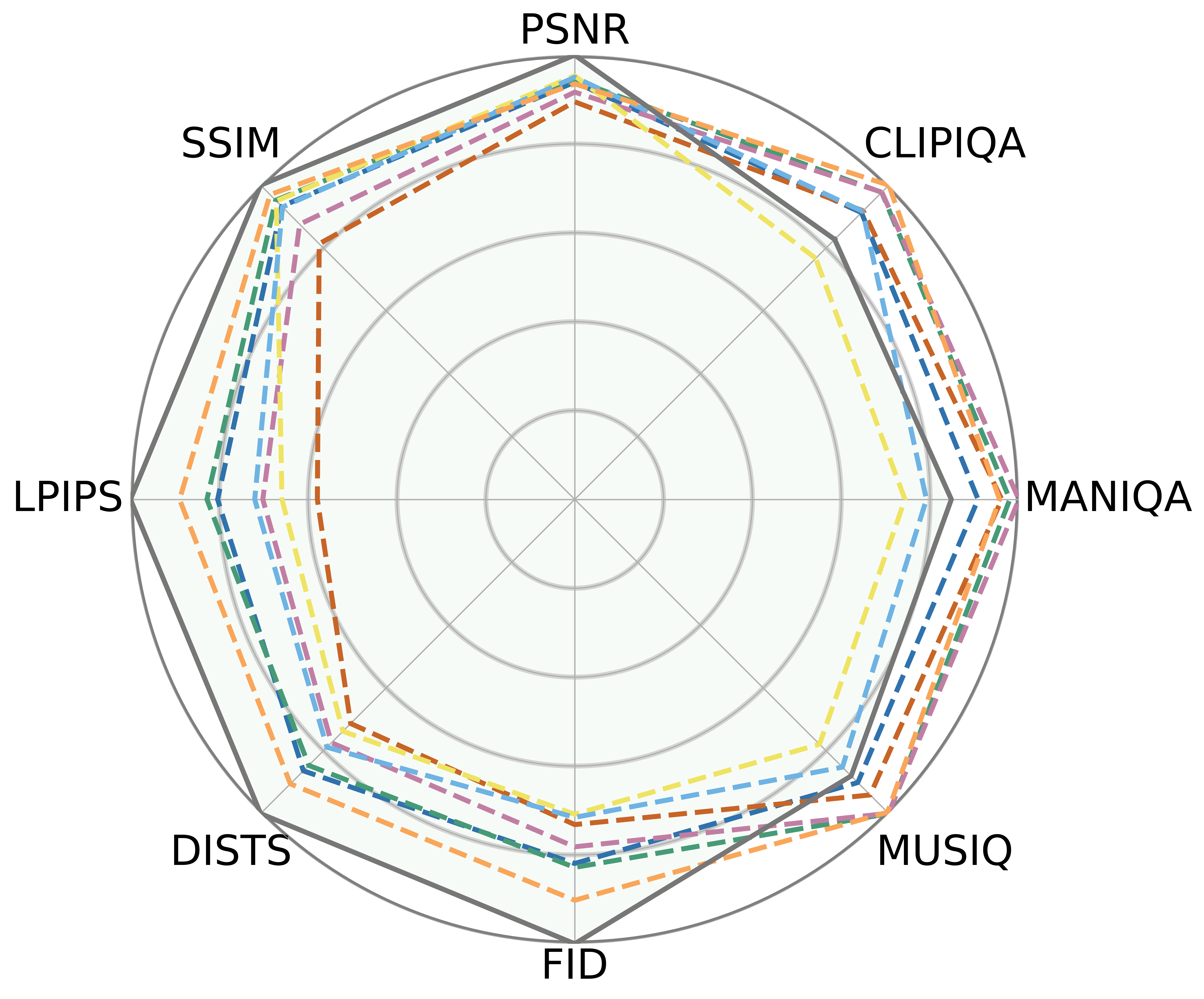}
        \caption{DRealSR}
    \end{subfigure}
    ~ 
    \begin{subfigure}[t]{0.26\textwidth}
        \centering
        \includegraphics[width=\textwidth,valign=m]{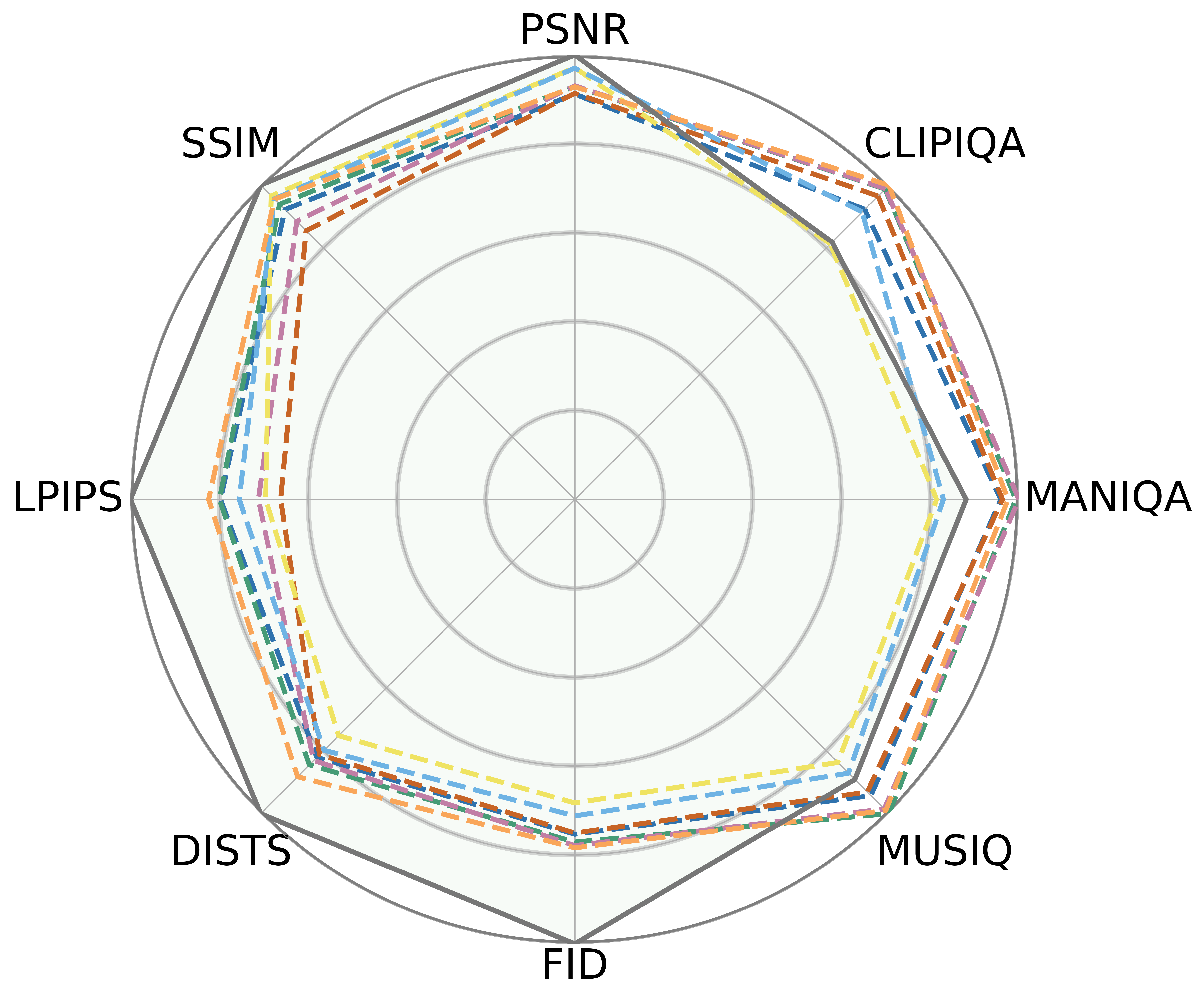}
        \caption{DIV2K}
    \end{subfigure}
    ~ 
    \begin{subfigure}[t]{0.14\textwidth}
        \centering
        \includegraphics[width=\textwidth,valign=m]{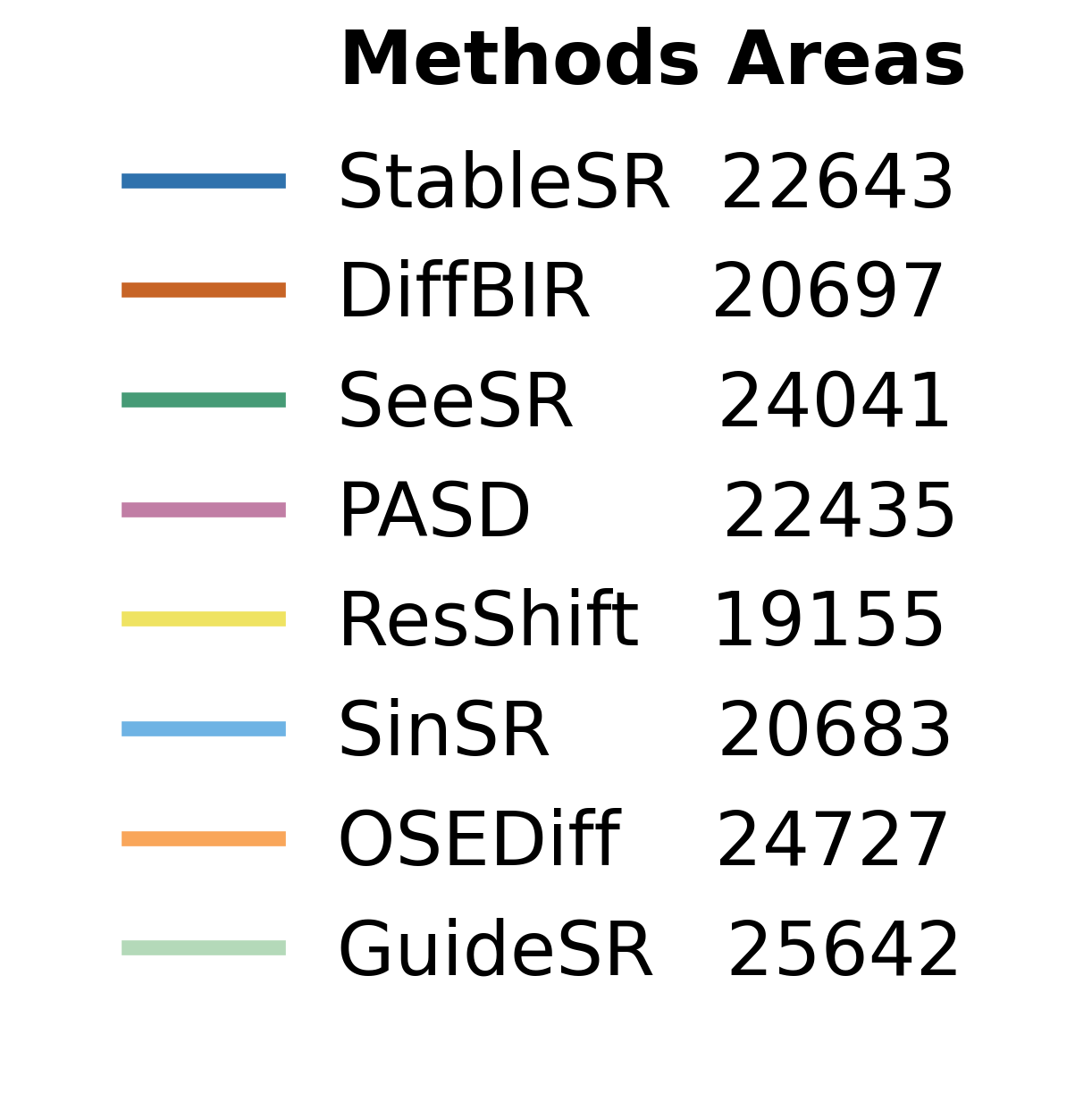}
    \end{subfigure}
    % \vspace{-3mm}
    % \caption{This radar chart visualizes the performance of various super-resolution methods across multiple evaluation metrics. The enclosed area for each method, computed using Cartesian transformations and the convex hull, provides a quantitative measure of overall effectiveness. A larger enclosed area indicates a more balanced and higher-performing method across all evaluated criteria. The computed areas are appended to the method names for direct comparison.}
    \caption{\sm{\textbf{Multi-metric performance visualization.} We visualize the performance of different SR methods, displaying reference-based metrics (PSNR, SSIM, LPIPS, DISTS, FID) on the left and no-reference metrics (CLIPIQA, MANIQA, MUSIQ) on the right. Our \xnet model prioritizes fidelity, consistently achieving the best reference-based metrics across all datasets. It does not lead in no-reference metrics due to the perception-distortion tradeoff~\cite{blau2018perception}. Despite this, \xnet consistently covers the largest area across all datasets, demonstrating a superior balance across various quality aspects. }}
    \label{fig:spider}
\end{figure*}

\noindent\textbf{Datasets.} Follow the training setup in ResShift~\cite{resshift}, we construct our training dataset using high-resolution (HR) images randomly cropped from ImageNet~\cite{imagenet} and FFHQ~\cite{ffhq}. The corresponding low-resolution (LR) images are synthesized using the degradation pipeline from Real-ESRGAN~\cite{realesrgan}, augmented with random horizontal flipping. The final HR target images are cropped to $256 \times 256$. For evaluation, we utilize the test sets provided by StableSR~\cite{stablesr}, which encompass both synthetic 3,000 $512\times 512$ images from DIV2K-val~\cite{div2k} with Real-ESRGAN degradation) and real-world images (paried LQ and $512\times 512$ HQ images from RealSR~\cite{realsr} and DRealSR~\cite{drealsr}). \smallskip

\noindent\textbf{Training Details.} We use a pretrained Stable Diffusion Turbo (v2.1) \cite{sauer2024fast} as our diffusion prior. RAM \cite{2023ram} is used as the prompt extractor. CLIP \cite{kumari2022ensembling} is used as the discriminator.
We train our model using the AdamW optimizer with an initial learning rate of $5 \times 10^{-5}$, following a cosine annealing schedule. To stabilize training, a warm-up phase of 500 iterations is applied, with a total iteration of 100k.  
All experiments are conducted on an 8-GPU cluster equipped with NVIDIA A100 GPUs, each with 40 GB of memory.

\begin{figure}
    \centering
    \includegraphics[width=0.8\linewidth]{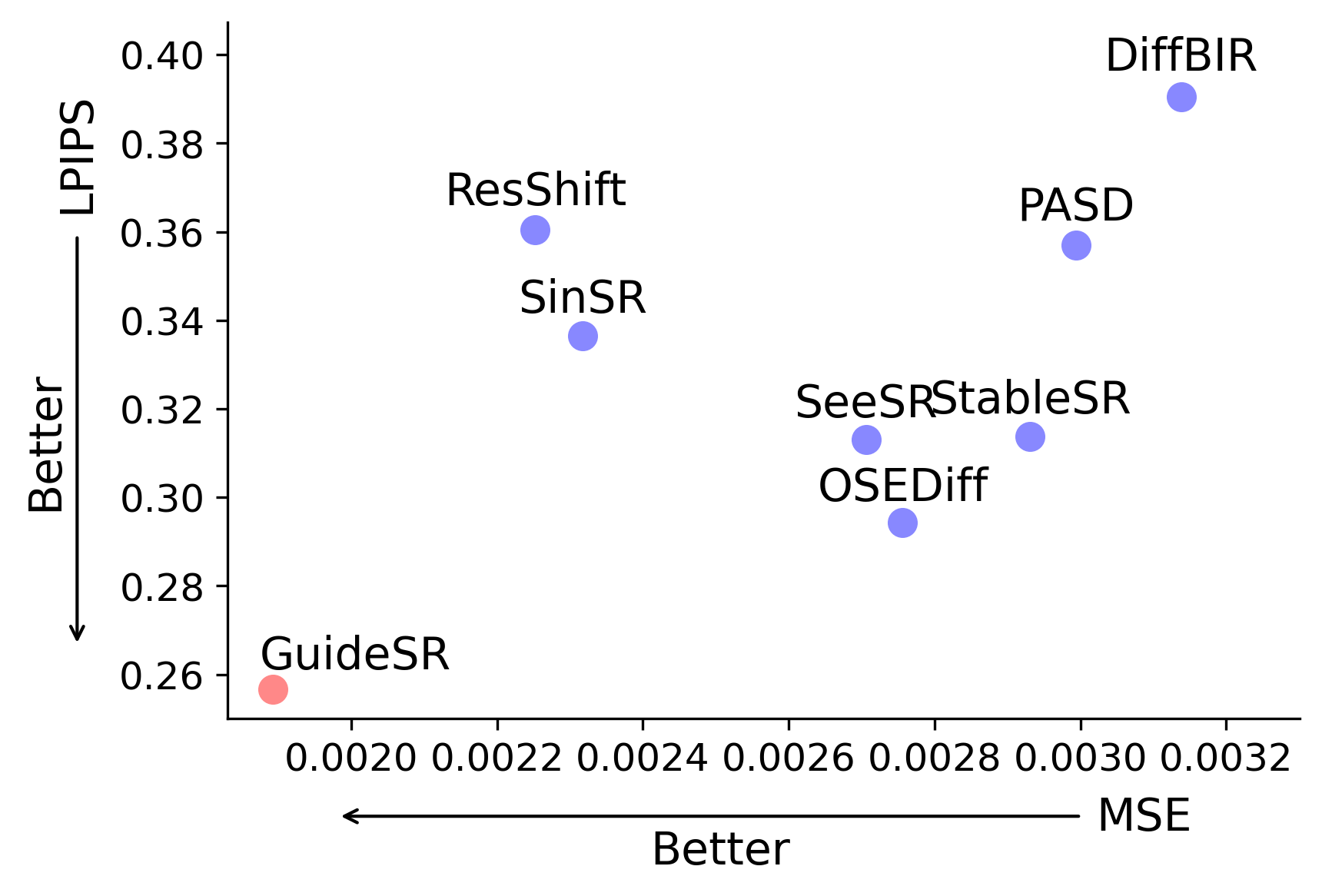}
    % \vspace{-3mm}
    % \caption{A comparative analysis of super-resolution methods using two reference-based evaluation metrics—Mean Squared Error (MSE) on the x-axis and Learned Perceptual Image Patch Similarity (LPIPS) on the y-axis. A method positioned closer to the bottom-left corner achieves both higher pixel-wise accuracy and better alignment with human perceptual preferences. \xnet (red) demonstrates the most favorable balance, achieving the lowest MSE while preserving structural and feature-level similarities. Other methods such as StableSR, PASD, and DiffBIR focus on perceptual alignment but at a cost of increased MSE.}
    \caption{\sm{\textbf{Pixel-Space Fidelity (MSE) and Feature-Space Fidelity (LPIPS).} Enhancing both MSE and LPIPS is generally challenging because boosting generative capabilities often increases feature-space fidelity and enhances the realism of restored images, but usually at the cost of reduced pixel-space fidelity. GuideSR achieves the highest scores in both MSE and LPIPS among all methods, demonstrating its ability to maintain fidelity in both pixel and feature spaces.}}
    \label{fig:mse-lpips}
    % \vspace{-8mm}
\end{figure}

\subsection{Quantitative Comparison}

We evaluate GuideSR against several state-of-the-art diffusion-based SR methods on three benchmark datasets: DIV2K-Val, DRealSR, and RealSR. As shown in Table~\ref{table:sota_table}, our method consistently outperforms both multi-step (StableSR~\cite{stablesr}, DiffBIR~\cite{diffbir}, SeeSR~\cite{seesr}, PASD~\cite{pasd}, and ResShift~\cite{resshift}) and single-step (SinSR~\cite{sinsr}, OSEDiff~\cite{osediff}) approaches across all reference-based metrics including PSNR, SSIM, LPIPS, DISTS and the distribution metric FID. \smallskip

\noindent\textbf{Quantitative Results on DIV2K-Val.} \sm{On the synthetic DIV2K-Val dataset, GuideSR achieves a PSNR of 24.76dB and an SSIM of 0.6333, surpassing the best previous method (ResShift) by 0.11dB and 0.0152 while requiring only a single inference step instead of 15. In terms of perceptual quality, our method demonstrates remarkable improvements with the lowest LPIPS (0.2653), DISTS (0.1879), and FID (21.04) scores among all methods. } \smallskip

\begin{figure*}[!t]
\begin{center}
\scalebox{0.76}{
\begin{tabular}[b]{c@{\hskip 8pt} c@{\hskip 8pt}  c@{\hskip 8pt} c@{\hskip 8pt} c@{ }}
% \hspace{-4mm}
    \multirow{4}{*}{\includegraphics[width=.4\textwidth,valign=t]{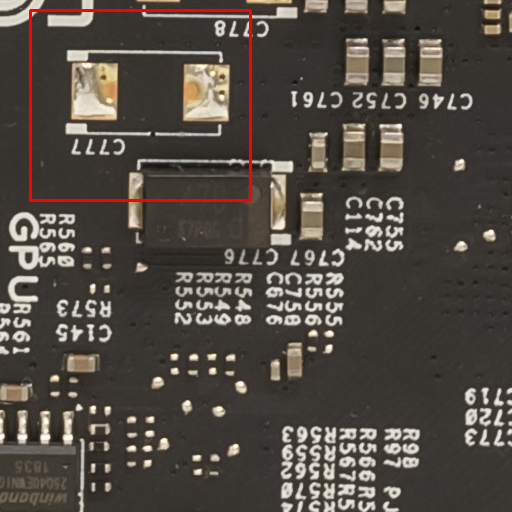}} & 
  	\includegraphics[width=.2\textwidth,valign=t]{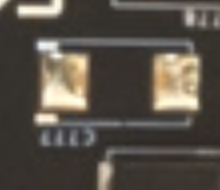}&   
    \includegraphics[width=.2\textwidth,valign=t]{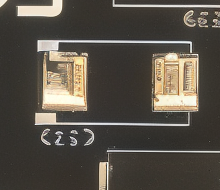}&
    \includegraphics[width=.2\textwidth,valign=t]{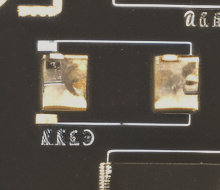}&
    \includegraphics[width=.2\textwidth,valign=t]{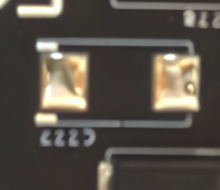}
\\
    &  Zoomed LQ &  DiffBIR \cite{diffbir}&  SeeSR \cite{seesr}&  PASD \cite{pasd} \\
     &
    \includegraphics[width=.2\textwidth,valign=t]{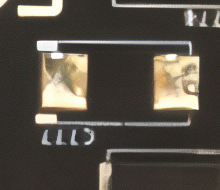}&  
     \includegraphics[width=.2\textwidth,valign=t]{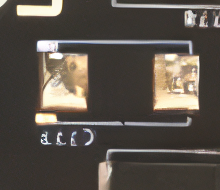}&
     \includegraphics[width=.2\textwidth,valign=t]{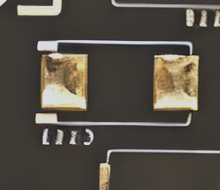}&
    \includegraphics[width=.2\textwidth,valign=t]{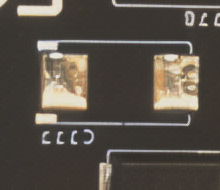}
     \\

               &  ResShift \cite{resshift} &  SinSR  \cite{sinsr}   &  OSEDiff \cite{osediff} &  \textbf{\xnet }
\\
\\
    \multirow{4}{*}{\includegraphics[width=.4\textwidth,valign=t]{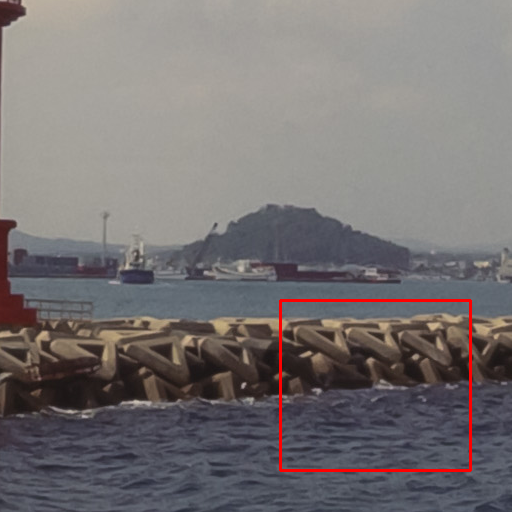}} &   
  	\includegraphics[width=.2\textwidth,valign=t]{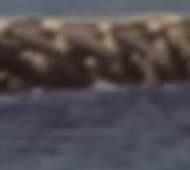}&   
    \includegraphics[width=.2\textwidth,valign=t]{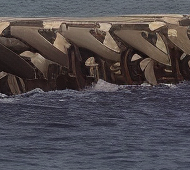}&
    \includegraphics[width=.2\textwidth,valign=t]{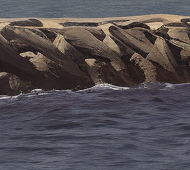}&
    \includegraphics[width=.2\textwidth,valign=t]{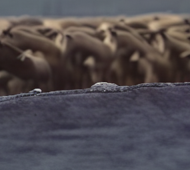}
\\
    &  Zoomed LQ &  DiffBIR \cite{diffbir}&  SeeSR \cite{seesr}&  PASD \cite{pasd} \\
     &
    \includegraphics[width=.2\textwidth,valign=t]{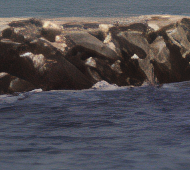}&  
     \includegraphics[width=.2\textwidth,valign=t]{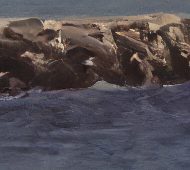}&
     \includegraphics[width=.2\textwidth,valign=t]{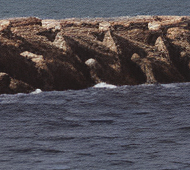}&
    \includegraphics[width=.2\textwidth,valign=t]{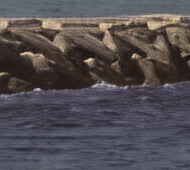}
     \\

               &  ResShift \cite{resshift} &  SinSR  \cite{sinsr}   &  OSEDiff \cite{osediff} &  \textbf{\xnet }
\\
\\
    \multirow{4}{*}{\includegraphics[trim={ 0 0 0 100
 },clip,width=.4\textwidth,valign=t]{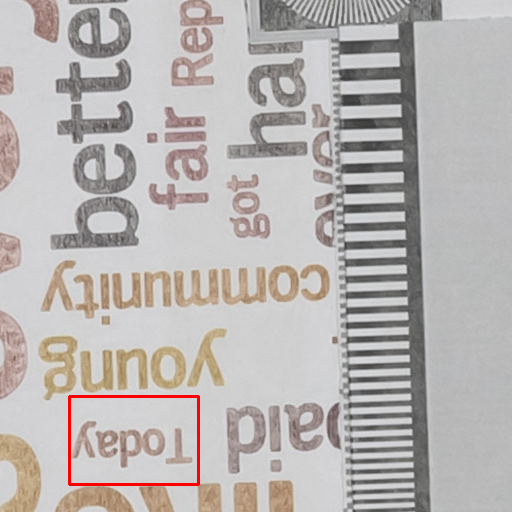}} &   
    \includegraphics[width=.2\textwidth,valign=t]{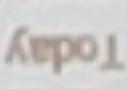}&   
    \includegraphics[width=.2\textwidth,valign=t]{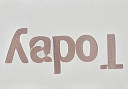}&
    \includegraphics[width=.2\textwidth,valign=t]{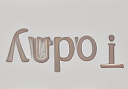}&
    \includegraphics[width=.2\textwidth,valign=t]{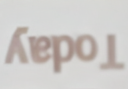}
\\
    &  Zoomed LQ &  DiffBIR \cite{diffbir}&  SeeSR \cite{seesr}&  PASD \cite{pasd} \\
     &
    \includegraphics[width=.2\textwidth,valign=t]{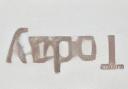}&  
     \includegraphics[width=.2\textwidth,valign=t]{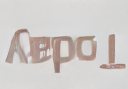}&
     \includegraphics[width=.2\textwidth,valign=t]{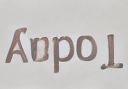}&
    \includegraphics[width=.2\textwidth,valign=t]{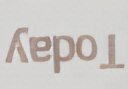}
     \\

               &  ResShift \cite{resshift} &  SinSR  \cite{sinsr}   &  OSEDiff \cite{osediff} &  \textbf{\xnet }
\\
\end{tabular}}
\end{center}
% \vspace{6mm}
% \caption{   \textbf{Image Super-Resolution on RealSR \cite{realsr}}. \xnet generates sharper and visually-faithful result.} 
\caption{\sm{\textbf{Visual Comparison on Real-World Images from RealSR~\cite{realsr} and DRealSR~\cite{drealsr}}.} (Top) \xnet accurately restores detailed features such as the text shape and the reflections on metal surfaces. (Middle) \xnet correctly reconstructs the geometry of the concrete blocks, while OSEDiff introduces incorrect textures and a color shift. (Bottom) \xnet faithfully restores the text, particularly the inverted ``a'' letter, whereas PASD generates authentic but slightly blurred text.}
\label{fig:realsr}
% \vspace{-0.5em}
\end{figure*}

\noindent\textbf{Quantitative Results on Real-World Datasets.} 
\sm{The advantages of GuideSR are even more pronounced on real-world datasets, which present more challenging and diverse degradations than synthetic ones. For example, on the DRealSR dataset, our method achieves a remarkable PSNR of 29.85dB, surpassing the best previous method (ResShift) by 1.39dB. The perceptual metrics also show significant improvements, with our method reducing the FID score to 122.06 on DRealSR (13.24 lower than the previous best) and 96.83 on RealSR (26.66 lower than the previous best). These consistent improvements across different real-world datasets demonstrate the robustness and generalizability of our approach.  } \smallskip

\noindent\textbf{Pixel-Space Fidelity and Feature-Space Fidelity. } Figure~\ref{fig:mse-lpips} visualizes the MSE and LPIPS of different methods on all three test sets. While MSE measures the fidelity in the pixel space, LPIPS measures the fidelity in the feature space and aligns well with human perception (perceptual similarity~\cite{lpips}). Achieving improvements in both MSE and LPIPS is typically challenging, as enhancing generative capabilities often leads to a decline in pixel-space fidelity while increasing the feature-space fidelity~\cite{resshift}. Nevertheless, \xnet achieves the best scores in both MSE and LPIPS among all methods, demonstrating its ability to preserve both pixel-space and feature-space fidelity. \smallskip

\noindent\textbf{Full-Reference and No-Reference Metrics.} Since \xnet focuses on the fidelity of the restored image, its performance is best evaluated by the full-reference metrics discussed above. That said, we also report no-reference image quality assessment (IQA) metrics including NIQE, MUSIQ, MANIQA and CLIPIQA for completeness. \xnet does not achieve the best performance on these metrics because of the perception-distortion tradeoff: Blau and Michaeli~\cite{blau2018perception} proved mathematically that less distortion (better full-reference scores\footnote{Notice that although feature-space metrics like LPIPS and DISTS highly correlate to human perception, they are full-reference metrics and are not considered perception metrics by definition~\cite{blau2018perception}.}) and better perceptual quality (better no-reference scores) cannot be achieved by the same image restoration algorithm. Given that \xnet achieves lowest distortion on all the three datasets across all full-reference metrics, it is unsurprising that it scores lower on no-reference metrics. Nevertheless, it performs comparably to existing methods. \smallskip

\noindent\textbf{Multi-Metric Performance Visualization.} 
To jointly compare the full-reference and no-reference metrics, Figure~\ref{fig:spider} visualizes the performance of different methods across all metrics using spider charts for each dataset. These radar plots provide an intuitive comparison  where larger area coverage indicates better overall performance across metrics. As shown in the figure, GuideSR consistently covers the largest area across all three datasets, achieving superior balance across different quality aspects. 
See the supplementary material for the details of this visualization.

\subsection{Qualitative Results}
Figure \ref{fig:realsr} presents visual comparisons between \xnet and other state-of-the-art methods on the RealSR dataset. \sm{The top row shows \xnet's ability to reconstruct electronic components with intricate patterns, capturing small details more accurately than other approaches, which often miss fine details or introduce false patterns.  
The middle row illustrates the reconstruction of triangular concrete blocks near a sea. While other methods struggle to accurately capture the exact block structure or unintentionally generate false textures, our method achieves most precise restoration of the complex geometry. OSEDiff also gives close geometry but introduces incorrect textures and a color shift.
The bottom row depicts the reconstruction of text. Although the proposed model is not specifically trained on text images, the guidance branch enables it to restore text with the greatest accuracy, especially for the inverted ``a'' letter. PASD also generates the correct letters but the result looks slightly blurry.} 
These qualitative results validate the effectiveness of our approach in preserving high-frequency details and structural integrity while enhancing perceptual quality, making GuideSR particularly suitable for real-world applications where both fidelity and visual quality are essential. \textbf{See the supplementary material for more qualitative results.}

\subsection{Ablation Study}
% \begin{figure}
%     \centering
%     \includegraphics[width=\linewidth]{example-image}
%     \caption{Ablation Table}
%     \label{fig:enter-label}
% \end{figure}

\begin{table}[]
    \centering
    \small
    \begin{tabular}{l|c}
    Experiment & PSNR $\uparrow$ \\
    \midrule
         Baseline & 26.65 \\
         Baseline + Long-skip & 26.80 \\
         Baseline + Guidance & 26.82 \\
         Baseline + Guidance + IGN + Long-skip & 27.08 \\
    \end{tabular}
    % \vspace{-2mm}
    \caption{Ablation study on RealSR~\cite{realsr} showing the contribution of different components to GuideSR's performance.}
    % Each row represents a different configuration, with PSNR values showing the incremental improvement as components are added.}    
    % Ablation Table - No experiment for Baseline + Guidance + IGN as IGN inherits a skip connection and it is counter intuitive to remove Long-skip for this experiment. No experiment for Baseline + IGN as the core facade is to pt enriched features in the middle Encode of DM. Passing through just IGN and putting that in Encoder is not helpful.}
    \label{table:ablation}
    % \vspace{-4mm}
\end{table}
To evaluate the contribution of each component, we conducted an ablation study using RealSR \cite{realsr}. The results are presented in Table~\ref{table:ablation}. \sm{``Baseline'' refers to using the Diffusion Branch only without the long-skip connections.} \smallskip

\noindent\textbf{Effect of Long-Skip Connections.} Adding long-skip connections to the baseline model improves the PSNR from 26.65dB to 26.80dB, demonstrating the importance of preserving low-frequency information throughout the network.  \smallskip

\noindent\textbf{Effect of Guidance Branch.} Incorporating the Guidance Branch provides a PSNR improvement from 26.65dB to 26.82dB over the baseline. This confirms our hypothesis that explicit structural guidance from the full-resolution input significantly enhances reconstruction quality. \smallskip

\noindent\textbf{Effect of Image Guidance Network (IGN).} The complete architecture (Baseline + Guidance + IGN + Long-skip) achieves the best performance with a PSNR of 27.08dB, which is 0.43dB higher than the baseline. This substantial improvement highlights the effectiveness of IGN in adaptively refining features and ensuring high-frequency details are preserved throughout the restoration process. \smallskip

\section{Conclusion}

In this study, we propose a super-resolution-centric framework designed to address the fidelity challenges present in existing one-step diffusion-based SR models. Our dual-branch architecture includes a Diffusion Branch with strong generative capabilities and a Guidance Branch that effectively extracts detailed structural and textural features. By utilizing the full-resolution degraded input as a guide, the Guidance Branch significantly enhances the fidelity of the reconstruction process while maintaining the computational efficiency of one-step SR models. \smallskip

\noindent\textbf{Limitations and future works.} Although one-step approaches significantly reduce inference time, the size of the denoising UNet still limits the checkpoint size and memory footprint, posing challenges for resource-constrained devices such as smartphones. Future work might focus on further improving efficiency through architectural optimization~\cite{kim2024bk, hu2024snapgen} and quantization~\cite{li2023q, huang2024tfmq}. Another potential direction is to extend the proposed framework to video restoration, where maintaining fidelity and computational efficiency is also crucial.

{\small
\bibliographystyle{ieeenat_fullname}
\bibliography{main}
}

\end{document}